# Stakeholder Perspectives on Digital Twin Implementation Challenges in Healthcare: Insights from a Provider Digital Twin Case Study

Md Doulotuzzaman Xames and Taylan G. Topcu
Grado Department of Industrial and Systems Engineering, Virginia Tech, Blacksburg, VA 24061, USA

## Abstract

Digital twin (DT) technology holds immense potential for transforming healthcare systems through real-time monitoring, predictive analysis, and agile interventions to support various decision-making needs. However, its successful implementation is contingent upon addressing an array of complex socio-technical challenges. Using a case study of provider workload DT, this research investigates DT implementation challenges in healthcare by capturing the perspectives of four distinct stakeholders: family medicine specialists (FMS), organizational psychologists (OP), engineers (EE), and implementation scientists (IS). We conducted semi-structured interviews guided by the updated Consolidated Framework for Implementation Research (CFIR 2.0), an implementation science framework widely used for understanding factors influencing implementation outcomes. We then mapped each stakeholder group's preferences and concerns, revealing a nuanced landscape of converging and diverging perspectives that highlight both shared and group-specific implementation barriers. Through thematic coding, the 66 identified challenges were categorized into seven domains: data-related, financial & economic, operational, organizational, personnel, regulatory & ethical, and technological. Our findings reveal shared concerns such as data privacy & security, interoperability, and regulatory compliance, highlighting shared priorities. However, divergences also emerged, reflecting each group's functional focus. For instance, while EEs emphasize technical issues like usability and scalability, FMSs prioritize practical challenges, including workload impact, staffing shortages, and resistance to change. OPs highlight collaboration, communication, and organizational alignment, while ISs concentrate on ethical issues and technology readiness. These findings emphasize the need for a multidisciplinary, stakeholder-sensitive approach that addresses both functional and practical concerns, highlighting the importance of tailored implementation strategies to facilitate successful DT adoption in healthcare.

## 1. Introduction

Digital twin (DT) technology has rapidly evolved into a powerful tool for enhancing system efficiency and decision-making across diverse industries, including manufacturing, aerospace, and energy [1]. A DT integrates a physical system with its virtual counterpart via a digital thread that enables continuous, bidirectional data/information exchange [2]. This dynamic connection allows for real-time monitoring, predictive analysis, and agile interventions, making DTs invaluable in managing complex systems [3].

Recently, DT applications in healthcare systems (HSs) have been rapidly growing, driven by the need to improve care quality, operational efficiency, and patient outcomes [4], [5]. Current research highlights DT use in four primary healthcare contexts: patient's body, medical procedures, healthcare facility, and public health [6]. DTs have demonstrated potential for early diagnosis, personalized treatment, and precision surgery, as well as for optimizing hospital workflows, staff scheduling, and resource allocation [7], [8]. However, despite their promise, most applications remain conceptual, and their practical integration faces significant barriers [6].

HSs are characterized by intricate sociotechnical interdependencies among workflows, technologies, and human actors, making implementation particularly complex [9], [10]. Effective DT implementation in healthcare depends on accurately identifying and navigating both technical and non-technical challenges [6]. Importantly, these challenges are often viewed differently by various stakeholders due to their distinct roles, expertise, and priorities. For example, healthcare providers may focus on the impact of DTs on clinician workload and staffing constraints, while engineers emphasize technological scalability and usability. Understanding and reconciling these differing perspectives are essential to developing comprehensive, stakeholder-sensitive strategies for successful DT adoption.

Despite its importance, research on stakeholder perspectives regarding DT implementation barriers in healthcare is nascent. To address this gap, this study presents findings from a DT implementation case study, by adopting the updated Consolidated Framework for Implementation Research (CFIR 2.0) [11] framework to extract preferences from key HS stakeholders in their unique operational context. Here, the use of CFIR is instrumental as it is a dedicated framework for investigating implementation outcomes in healthcare; thus, it offers a structured approach to understanding the sociotechnical complexities inherent in adopting new technologies [12]. Guided by CFIR 2.0, we conducted semi-structured interviews with stakeholders from four distinct groups – family medicine specialists (FMS), organizational psychologists (OP), engineers (EE), and implementation scientists (IS) – to systematically capture their concerns and preferences across the framework's five domains. This mapping reveals insights into shared and group-specific barriers, offering a comprehensive understanding of the factors influencing DT implementation in healthcare. The central research question guiding this investigation is: *"How do the perspectives of different stakeholders converge or diverge in identifying and prioritizing challenges related to the implementation of DTs in HSs?"*

## 2. Methods

This study adopts a qualitative approach guided by the CFIR 2.0 framework to investigate the challenges of implementing DTs in HSs, summarized in Figure 1. Below we briefly discuss each step.

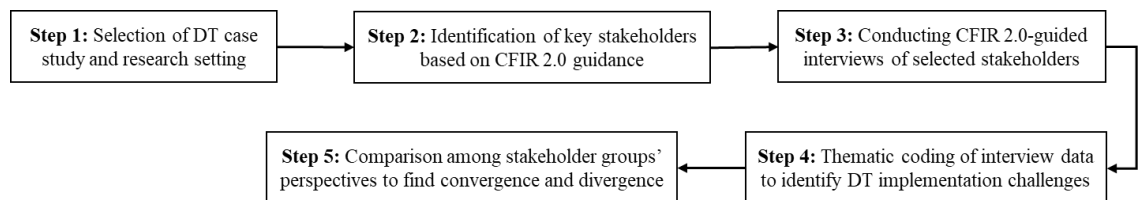

Figure 1: Methodology adopted in this study.

- In Step 1, we selected the case study that involves the development and implementation of a DT for managing healthcare provider workload, with a long-term objective of mitigating burnout [13]. The case study took place in a family medicine clinic, a primary care facility with over 50 practicing physicians.
- In Step 2, following CFIR 2.0 guidance, four key stakeholder groups were identified to capture perspectives on DT implementation challenges: FMS, OP, EE, and IS. Purposive sampling was used to select eight stakeholders (two from each group) with substantial experience in their respective fields.
- In Step 3, semi-structured interviews were conducted with stakeholders using an interview protocol aligned with the 39 constructs of CFIR 2.0. This framework facilitated a structured yet flexible approach, allowing for the exploration of both shared and role-specific challenges.
- In Step 4, the interview data were analyzed using thematic coding techniques. Preliminary findings were shared with participants to validate the accuracy and relevance of interpretations.
- In Step 5, we mapped the identified themes across stakeholder groups to highlight areas of convergence and divergence. Convergence refers to shared concerns that emphasize common implementation priorities, while divergence highlights role-specific differences. This comparison provided a deeper understanding of implementation barriers and informed strategies for addressing both functional and practical concerns.

## 3. Findings

Our CFIR 2.0-guided stakeholder interviews identified 66 implementation challenges, which we categorized into seven groups through thematic coding: data-related (n=7), financial & economic (n=5), operational (n=8), organizational (n=18), personnel (n=12), regulatory & ethical (n=3), and technological (n=13). Due to the extensive width of the table, we divided it into two: Table 1 for the first four categories, and Table 2 for the other three.

The first row of both tables highlights challenges unanimously identified by all stakeholder groups, including data privacy, data security, collaboration and communication barriers, and ethical concerns. These challenges demand the most urgent attention in DT implementation in healthcare. Rows 2, 3, and 4 of Tables 1 and 2 display challenges identified by three stakeholder groups, highlighting areas of convergence and shared concerns in the implementation process. For instance, challenges such as validation & verification, interoperability, and testing & evaluation were highlighted by EE, FMS, and IS, underscoring the collective recognition of these critical issues. Similarly, organizational inertia/resistance to change, the organization's technology readiness, and high initial costs were identified by EE, IS, and OP, reinforcing their significance across different perspectives.

Table 1: A distribution of the shared and unique challenges identified by different stakeholders in four categories

| Stakeholder group | Data-related challenges | Financial & Economic challenges | Operational challenges | Organizational challenges |
|---|---|---|---|---|
| EE, FMS, IS, OP | Data privacy; Data security | Unclear/intangible benefits | - | Collaboration and communication barriers; Identification of the need/problem |
| EE, FMS, IS | High real-time data needs | - | Scope management in a dynamic landscape | - |
| EE, FMS, OP | Data accessibility | Willingness for upfront investment | - | - |
| EE, IS, OP | - | High initial cost/Financial viability concerns | - | Organizational inertia/resistance to change; Organization's technology readiness |
| EE, IS | - | - | Lack of generalizability | - |
| EE, OP | - | - | Lack of training and support infrastructure | - |
| FMS, IS | - | - | - | Misalignment of outcomes and incentives; Effective reflection, evaluation, and feedback mechanisms |
| FMS, OP | - | - | Quantification of benefits and outcomes | Alignment with organizational goals and stakeholder perceptions/Organization prioritizing other outcomes; Pressure for immediate operational performance |
| IS, OP | - | - | - | Stakeholder engagement |
| EE | Data integration/fusion | - | - | Key performance indicators (KPIs) tracking and management |
| FMS | Data ownership | ROI uncertainty; Lack of immediate benefits | Need for integrated decision support systems; Availability of resources to implement operational changes. | Resilience to low-frequency, high-impact disruptions; Free trialability; Cultural emphasis on reactive work and reporting over foundational system issues; Differences in leadership styles; Presence of shadow influencers; Change fatigue |
| IS | - | - | Trust and transparency in implementation | - |
| OP | Data governance | - | Reversibility of operational decisions | Lack of established best practice guidelines; Value communication |

Additionally, Tables 1 and 2 also present challenges acknowledged by two stakeholder groups, as well as those unique to specific groups. For example, both FMS and OP identified challenges related to alignment with organizational goals, stakeholder perceptions, and pressure for immediate operational performance under the organizational challenges category, illustrating areas where concerns overlap between distinct stakeholders.

Table 2: A distribution of the shared and unique challenges identified by different stakeholders in three categories

| Stakeholder group | Personnel challenges | Regulatory & Ethical challenges | Technological challenges |
|---|---|---|---|
| EE, FMS, IS, OP | Trust in DT developers/vendors | Ethical issues | - |
| EE, FMS, IS | Individual's inertia/resistance to change; Individual's commitment to implementation | Legal issues; Regulatory compliance | Validation and verification; Interoperability/Integration complexity; Testing and evaluation |
| EE, FMS, OP | Lack of prior research/Lack of evidence in human-in-the-loop systems; Providers' perception on the technology/Technophobia among older generation providers | - | - |
| EE, IS | Lack of understanding of the technology | - | - |
| FMS, IS | Differences in motivation levels among individuals/groups | - | - |
| EE | - | - | Lack of standardization; User-specific personalization; Scalability; Usability issues |
| FMS | Staffing shortage; Fear of added workload; Lack of expertise; Lack of "understanding" of perceived benefits; Provider schedule constraints for technology learning | - | Human-work interaction design; Time-scale dependency of DT performance; Model latency & timeliness; Performance concerns; Data fragmentation/Siloed data sources |
| OP | - | - | Infrastructure and workflow integration |

Figure 2 presents a heatmap that visualizes the distribution and intensity of challenges reported by different stakeholder groups in the DT implementation study. The heatmap quantifies the number of challenges identified within each category, with darker shades indicating a higher concentration of reported challenges. This visualization provides a structured comparison of stakeholder perspectives, revealing both overlapping and distinct concerns. By mapping these concerns, we gain insight into how different professional backgrounds influence perceptions of DT implementation challenges and where targeted interventions may be necessary to address critical adoption barriers.

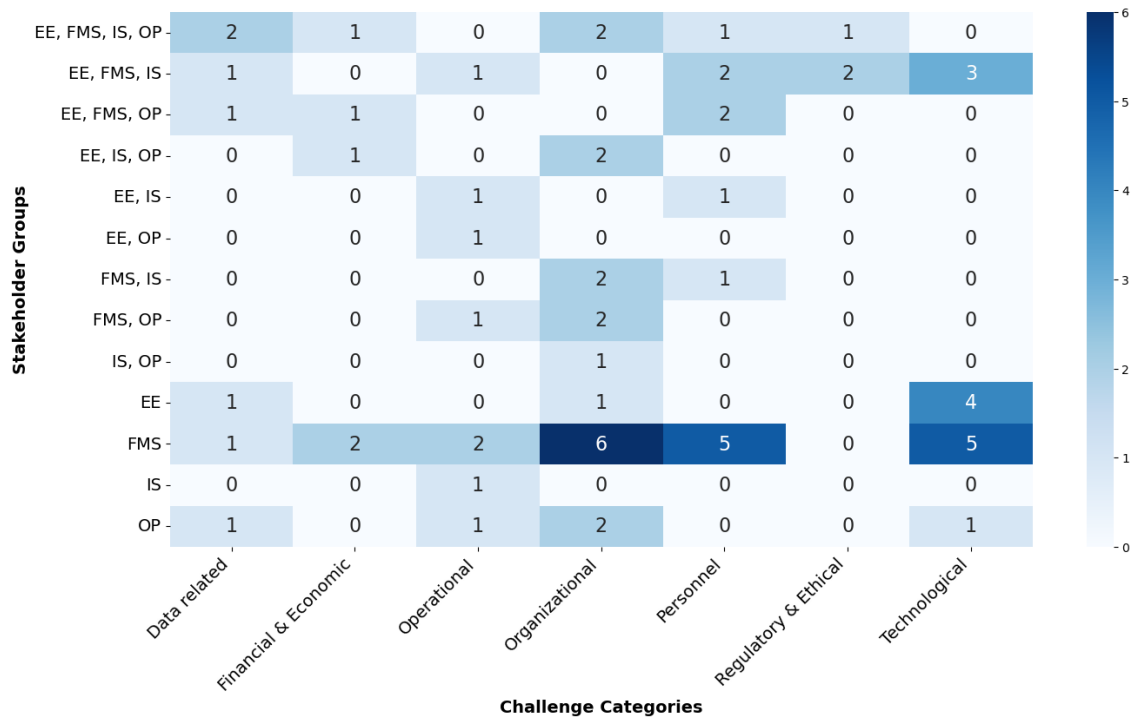

Figure 2: Heatmap of DT implementation challenges identified by different stakeholder groups.

Despite differences in expertise and roles, all stakeholder groups unanimously identified regulatory and ethical challenges as a central concern, emphasizing a shared prioritization of compliance, patient safety, and governance in DT adoption. This widespread agreement underscores the reality that DTs, despite their potential to enhance healthcare operations, introduce risks that must be carefully managed to maintain legal and ethical integrity. The convergence on data privacy, interoperability, and collaboration barriers further reinforces the necessity for robust security frameworks, seamless system integration, and interdisciplinary cooperation. These findings suggest that certain foundational barriers are universal and must be proactively addressed for DT implementation to be both technically viable and clinically acceptable. Failure to do so could lead to adoption hesitancy, operational inefficiencies, and potential regulatory violations, ultimately limiting the scalability of DT applications.

Beyond these shared concerns, the heatmap reveals significant divergences in how stakeholder groups perceive the scope and impact of DT challenges, reflecting their distinct functional priorities. As shown in Figure 2, FMS reported the highest number of unique challenges, primarily in the organizational and personnel domains. Their concerns about workload burden, staffing shortages, and resistance to change suggest that, beyond their clinical roles, FMS are deeply engaged in the operational realities of DT adoption. This finding highlights the critical role of frontline healthcare providers in assessing the feasibility of new technologies, as their daily workflows are directly impacted by DT-driven decision support systems. Without addressing the organizational and workforce-related challenges that FMS emphasize, DT adoption could exacerbate clinician burnout, further straining an already overburdened healthcare workforce.

In contrast, EE identified the highest concentration of technological challenges, focusing on interoperability, validation, and scalability. This focus aligns with their responsibility for system architecture, integration, and performance optimization. Their concerns highlight the complexities involved in ensuring DT systems can seamlessly interface with existing healthcare IT infrastructure, an essential requirement for real-world deployment. The prominence of validation and verification concerns further emphasizes the need for rigorous testing frameworks to ensure DT predictions are both reliable and clinically actionable. These findings indicate that technical readiness remains a critical barrier to DT implementation, requiring ongoing refinement of algorithms, data pipelines, and user interfaces to meet healthcare standards.

Meanwhile, OP emphasized challenges related to stakeholder engagement, collaboration, and organizational readiness, underscoring the human and organizational factors that influence technology adoption. Their concerns suggest that successful DT implementation is not just a technical challenge but also an issue of organizational change management. Resistance from healthcare professionals, unclear communication of DT benefits, and a lack of structured change facilitation mechanisms can all impede adoption. These findings suggest that interdisciplinary implementation strategies – integrating behavioral insights with technical solutions – are critical for aligning stakeholder expectations and ensuring smoother transitions to DT-driven workflows.

IS, while identifying fewer unique challenges, focused on regulatory compliance, trust in DT vendors, and ethical considerations. Their perspective offers a macro-level view of systemic barriers, emphasizing that even a technically sound DT solution will fail without proper regulatory alignment and stakeholder trust. The concerns raised by IS reflect broader governance issues, including the need for standardized DT evaluation metrics, transparent accountability structures, and mechanisms to ensure equitable technology deployment. Addressing these issues is crucial, as failure to establish clear legal and ethical guidelines could hinder the widespread adoption of DTs in healthcare.

## 4. Discussion and Conclusions

HSs invest significant financial, technical, and clinical resources in interventions (e.g., new technologies, processes, guidelines) to improve care quality and efficiency [14]. However, even high-quality, evidence-based technologies often fail due to misalignment with daily workflows and operational realities [15]. DTs are emerging rapidly, promising real-time decision support and predictive capabilities, yet little is known about how key HS stakeholders perceive DT implementation challenges. To address this gap, we leveraged a provider workload DT case study engaging four stakeholder groups: FMS, EE, OP, and IS. Our goal was to uncover shared and stakeholder-specific challenges influencing DT adoption.

Findings reveal strong convergence on fundamental concerns such as data privacy, regulatory compliance, and communication barriers, underscoring universal priorities that transcend disciplinary boundaries. Failures in these areas can compromise patient trust, safety, and system reliability. Widespread agreement on regulatory and ethical challenges highlights concerns about governance and accountability in DT-enabled decision-making. Similarly, data privacy and interoperability emerged as critical challenges, reinforcing the need for robust security frameworks and seamless data integration to prevent DT adoption from becoming a liability.

Beyond shared concerns, each stakeholder group exhibited distinct priorities. FMS, directly engaged in patient care, identified the highest number of challenges, particularly within organizational and personnel domains, emphasizing workload, staffing shortages, and provider resistance. This suggests that, beyond clinical responsibilities, FMS also navigate operational constraints affecting DT feasibility. EE focused on technological concerns such as validation, scalability, and integration complexity, reflecting their responsibility for ensuring DT functionality. OP emphasized stakeholder engagement and institutional readiness, recognizing that even the most advanced technology will fail without organizational buy-in. IS contributed fewer unique challenges but provided a system-wide perspective, focusing on regulatory, ethical, and economic concerns affecting long-term viability.

These perspectives illustrate the multifaceted nature of DT implementation challenges. While some barriers require standardized solutions, others are highly contextual, necessitating tailored strategies for different stakeholders. Failure to navigate this complexity thoughtfully risks repeating mistakes seen with electronic health records (EHR) systems, where inadequate stakeholder alignment led to inefficiencies and clinician burnout [16]. To avoid this, DT developers must balance technological advancement with frontline usability, ensuring implementation strategies reflect the realities of those using these systems.

A key insight from this research is the complexity introduced by sociotechnical interactions within HSs. Addressing technological issues in isolation is insufficient; organizational and personnel-related factors are equally critical.

Organizational inertia, lack of training infrastructure, and misaligned incentives, as identified by multiple stakeholders, suggest that healthcare organizations must prioritize fostering a culture of change, improving workforce readiness, and aligning operational goals with DT implementation.

In conclusion, successful DT implementation requires an integrated, stakeholder-sensitive approach that balances technical innovation with organizational adaptability and personnel engagement. This study underscores the importance of multidisciplinary collaboration in overcoming the complex barriers identified by stakeholders. By addressing both shared and role-specific concerns, HSs can unlock the transformative potential of DTs for improved operational efficiency and patient care. Future research should develop practical guidelines for tailored DT adoption strategies to systematically address implementation challenges in healthcare.

## Acknowledgment

The authors sincerely appreciate the support received from Virginia Tech's Institute for Society, Culture, and Environment (ISCE) and the Destination Areas 2.0 Planning and Development Grant. They also extend their heartfelt thanks to the interview participants for generously sharing their time and insights, which were instrumental in conducting this research.